\documentclass[twocolumn,showpacs,preprintnumbers,amsmath,amssymb,floatfix]{revtex4}

\usepackage{graphicx}
\usepackage{dcolumn}
\usepackage{bm}

\usepackage[normalem]{ulem}  
\usepackage[dvips]{color}

\begin{document}


\title{Probing the symmetry energy from the nuclear isoscaling}

\author{S.R.\ Souza$^{1,2}$}
\author{M.B.\ Tsang$^3$}
\author{R.\ Donangelo$^1$}
\author{W.G.\ Lynch$^3$}
\author{A.W.\ Steiner$^3$}
\affiliation{$^1$Instituto de F\'\i sica, Universidade Federal do Rio de Janeiro
Cidade Universit\'aria, \\CP 68528, 21941-972, Rio de Janeiro, Brazil}
\affiliation{$^2$Instituto de F\'\i sica, Universidade Federal do Rio Grande do Sul\\
Av. Bento Gon\c calves 9500, CP 15051, 91501-970, Porto Alegre, Brazil}
\affiliation{$^3$ Joint Institute for Nuclear Astrophysics, National
Superconducting Cyclotron Laboratory, and the Department of Physics
and Astronomy, Michigan State University,
East Lansing, MI 48824, USA}

\date{\today}

\begin{abstract}
Using different parameterizations of the nuclear mass formula, we study
the sensitivity of the isoscaling parameters to the mass formula
employed in grand-canonical calculations.
Previous works on isoscaling have suggested that the symmetry energy implied
in such calculations is anomalously smaller than that suggested by fits to
nuclear masses.
We show that surface corrections to the symmetry energy naturally broadens the
isotopic distribution thus allowing for values of the symmetry energy
which more closely match those obtained from nuclear masses.
\end{abstract}

\maketitle

\section{\label{sec:introuduction}Introduction\protect}

Understanding multi-fragmentation,  the breakup of nuclei into many
fragments, has challenged theorists and experimentalists for almost
three decades
\cite{Jakobsson1982,Finn1982,Moretto1993,reviewSubal2001,BettyPhysRep2005}.
This phenomenon can be observed in both relativistic light-ion
collisions and in central nucleus-nucleus collisions, even though
the role of reaction dynamics in these two domains can be quite
different. The consequential variation in the mixture of statistical
and time-dependent phenomena in multi-fragmentation complicates the
application of the traditional theoretical tools. Nonetheless,
statistical techniques have been successfully applied to calculate
the relative production of various fragments
\cite{reviewSubal2001,BettyPhysRep2005,Moretto1993,Bondorf1995,ismmFlow2007},
even though the fundamental statistical parameters such as
temperature or density are not external constraints, but must be
inferred from data (see
\cite{reviewSubal2001,BettyPhysRep2005,ThermometryLeFevre,ThermometryNatowitz}
and references therein).

It is therefore natural to apply statistical methods to fragment
isotope distributions and to their isoscaling
behavior \cite{isoscaling1,isoscaling2}. The latter term refers to
the scaling properties observed in Refs.\
\cite{isoscaling1,isoscaling2}, that govern the experimentally
observed ratios between the yields $Y_i(N,Z)$ associated with an
emitted fragment of atomic and neutron numbers $Z$ and $N$,
respectively, produced in two similar reactions, labeled $i=1,2$:

\begin{equation}
R_{21}(N,Z)=Y_2(N,Z)/Y_1(N,Z)\propto \exp(\alpha N + \beta Z)\;,
\label{eq:isoscaling}
\end{equation}

\noindent where $\alpha$, and $\beta$ are the isoscaling parameters.
The isoscaling of fragment isotopic yields has also been observed in
deep inelastic collisions, evaporation processes, as well as in
nuclear multifragmentation reactions
\cite{isoscaling2,isoYennello2003,isoSouliotis2003}. It has been
exploited by many authors to obtain information on observables
related to the multifragment emission
\cite{isoscaling2,isoscWolfgangBotvina1,RadutaIsoSym,isocc,isoscalingIndraGSI2005}.
In particular, a link between the symmetry energy coefficient
$C_{\rm sym}$ and the isoscaling parameter $\alpha$ has been
suggested in Refs.\ \cite{isoscaling2,isoscWolfgangBotvina1},

\begin{equation}
\alpha=\frac{4C_{\rm sym}}{T}
        \left(\frac{Z_1^2}{A_1^2}-\frac{Z_2^2}{A_2^2}\right)\;.
\label{eq:alphaGamma}
\end{equation}

\noindent If a reaction allows straight-forward application of
statistical approaches, it may be possible to obtain empirical
constraints on $C_{\rm sym}$, at different values of the breakup
density and temperature $T$, from the knowledge of the decaying
sources $(A_1,Z_1)$ and $(A_2,Z_2)$ or, the proton fractions,
$Z_1/A_1$ and $Z_2/A_2$, together with $\alpha$ and $T$. To explore
this possibility, many studies along this line have been carried
out, which rely directly \cite{isoscalingIndraGSI2005} or indirectly
\cite{isoscWolfgangBotvina1,isoSymmetryBotvina2006,symEnergyShatty2007,isoNatowitz2007,isoNatowitz2007_2}
on the close connection between the symmetry energy and the
isoscaling parameter $\alpha$. Such studies require explicit
assumptions about the form for the symmetry free energy and about
how the symmetry free energy depends on temperature.

In this work, we investigate these assumptions. We explore the
sensitivity of $\alpha$ to the mass formula employed in statistical
calculations. In addition to the standard parametrization of the
binding energy~\cite{isoscWolfgangBotvina1}, we also consider mass
formulae in which the surface effects in the symmetry energy are
included \cite{PawelMasFormula,ISMMmass}. We examine the rigor of
present extractions of the temperature dependence of the symmetry
energy from isoscaling analyses and whether such analyses presently
provide unambiguous information on this temperature dependence. In
sect.\ \ref{sec:model}, we present the statistical treatment
employed in this work, as well as the mass formulae we use. The
sensitivity of the isoscaling parameters to the binding energy is
then investigated in sect.\ \ref{sec:results}. Our main conclusions
are delineated in sect.\ \ref{sec:conclusions}.

\section{\label{sec:model}The model\protect}
The Statistical Multifragmentation Model (SMM) \cite{smm1,smm2,smm4}
is based on a scenario in which a hot source with mass and atomic
numbers $A_0$ and $Z_0$, respectively, at temperature $T$, expands
to a density $\rho=\rho_0/(1+\chi)$ and undergoes a prompt breakup.
The normal nuclear density is denoted by $\rho_0$ and $1/\chi$
corresponds to the ratio between the breakup volume and that
available to the fragments. The properties of the system are
calculated at the freeze-out stage, where it is explicitly assumed
that, except for the Coulomb repulsion, there is no interaction
among the fragments. Some interactions of the fragments with the
surrounding light particle gas are incorporated via the temperature
dependence of the nuclear surface energy. Within the SMM approach,
the total Helmholtz free energy of the system is approximated by

\begin{eqnarray}
F(T,V) &=& n_1 (f_1+f_1^{\rm trans})+\cdots
          +n_M (f_M+f_M^{\rm trans})
\nonumber\\
  &+& \frac{C_{\rm c}}{(1+\chi)^{1/3}}
\frac{\left(\sum_i n_i Z_i\right)^2}
{\left(\sum_i n_i A_i \right)^{1/3}}\;,
\label{eq:feg}
\end{eqnarray}

\noindent
where $n_i$ denotes the multiplicity of species $i$, $f_i$ and
$f_i^{\rm trans}$ respectively stand for the internal and translational
contributions of a fragment of species $i$ to the free energy.
The last term in the above expression corresponds to the homogeneously
charged sphere which, with the other terms contained in $f_i$, give the
Wigner-Seitz approximation to the Coulomb interaction among the
fragments \cite{WignerSeitz,smm1}.
The coefficient $C_c$ is taken from a mass formula (see below).
The species multiplicities are constrained by the properties of the decaying
source:

\begin{eqnarray}
\label{eq:constA}
A_0=\sum_i^MA_i n_i\\
Z_0=\sum_i^MZ_i n_i\;.
\label{eq:constZ}
\end{eqnarray}

\subsection{\label{sec:gc}The grand canonical approach}
For clarity, we use the grand-canonical version of SMM, from which
analytical expressions for the average fragment multiplicities can
be easily obtained.

By construction, the terms involving $f_i(T,V)$ have no explicit or
implicit dependence on $n_i$, since it has contributions only from the
binding energy $B_i$, the internal excitation energy of the fragment and
the remaining Wigner-Seitz Coulomb terms of Eq.\ (\ref{eq:feg}).
Thus, at this point, one only needs the explicit expression for
$f_i^{\rm trans}$ which, through the Stirling's formula,
$\log n!\approx n\log n-n$, can be written as

\begin{equation}
f_i^{\rm trans}(T,V,n_i)=
-T\left[\log\left(\frac{g_iV_f}{\lambda_T^3}A^{3/2}_i
\right)-\log n_i+1\right]
\label{eq:fe_trans}
\end{equation}

\noindent
where $\lambda_T=\sqrt{2\pi\hbar^2/mT}$, $m$ is the nucleon mass,
$g_i$ denotes the spin degeneracy factor, and $V_f$ stands for the
free volume.

The average fragment distribution may be obtained by minimizing the total
free energy of the system with respect to $\{n_k\}$, subject to the
constraints (\ref{eq:constA})-(\ref{eq:constZ}):

\begin{equation}
\frac{\partial {\cal F}}{\partial n_k}=0\;,
\label{eq:dfzero}
\end{equation}

\noindent
where

\begin{equation}
{\cal F}=F+\mu_B\left[A_0-\sum_iA_i n_i\right]
       +\mu_Q\left[Z_0-\sum_iZ_i n_i\right]\;,
\label{eq:fconst}
\end{equation}

\noindent
and $\mu_Q$ and $\mu_B$ are the Lagrange multipliers, which turn out to be
the charge and baryon chemical potentials, respectively.
Thus, the above relations lead to

\begin{eqnarray}
0&=&f_k-T\left[\log\left(\frac{g_iV_f}{\lambda_T^3}A^{3/2}_i
\right)-\log n_i+1\right]+T
\nonumber\\
&+&\frac{2Z_kC_{\rm c}}{(1+\chi)^{1/3}}
\frac{\sum_i n_i Z_i}{\left(\sum_i n_i A_i\right)^{1/3}}\nonumber\\
&-&\frac{A_k C_{\rm c}}{3(1+\chi)^{1/3}}
\frac{\left(\sum_i n_i Z_i\right)^2}{\left(\sum_i n_i A_i\right)^{4/3}}
-\mu_BA_k-\mu_QZ_k\;.
\label{eq:fezero}
\end{eqnarray}

\noindent
Upon using Eqs. (\ref{eq:constA})-(\ref{eq:constZ}), one finally arrives at

\begin{eqnarray}
n_k &=& \frac{g_k V_f A_k^{3/2}}{\lambda_T^3}
\exp\Big\{-\Big(f_k-\mu_B A_k-\mu_Q Z_k\nonumber\\
&+&\frac{2Z_kZ_0C_{\rm c}}{A_0^{1/3}(1+\chi)^{1/3}}
-\frac{A_kZ_0^2C_{\rm c}}{3A_0^{4/3}(1+\chi)^{1/3}}\Big)/T
\Big\}\;.
\label{eq:aven}
\end{eqnarray}

\noindent
It should be noticed that the last two terms in the above expression
arise from the homogeneous term of the Wigner-Seitz approximation.
Since the chemical potentials can be redefined to absorb these terms,
they do not play a role in the determination of the particle
multiplicity. However, they modify the relation between $\alpha$ and
the mass formula, inasmuch as the difference between the chemical
potentials in reactions 1 and 2 is the relevant quantity to the
calculation \cite{isoscaling2,isoscWolfgangBotvina1}. Nevertheless,
since the considered sources are usually similar, the symmetry energy
still gives the main contribution to $\alpha$ and, therefore, this
correction is neglected.

The internal free energy of the species $i$ is given by

\begin{eqnarray}
f_i &=&-B_i-\frac{T^2}{\epsilon_0}A_i
   +\beta_0A^{2/3}_i\left[\left(\frac{T_c^2-T^2}{T_c^2+T^2}\right)^{5/4}-1\right]
\nonumber\\
&-&C_{\rm c}\frac{Z_i^2}{A_i^{1/3}}\frac{1}{(1+\chi)^{1/3}}\;,
\label{eq:fi}
\end{eqnarray}

\noindent
where $B_i$ denotes the binding energy of the fragment,
$\epsilon_0=16$~MeV, $\beta_0=18$~MeV and $T_c=18$~MeV. The last term
in the above relation is associated with the Wigner-Seitz correction,
whereas the Coulomb self energy of the fragment is already included in
$B_i$.
The form of the suppression of the surface energy as $T
\rightarrow T_C$ was determined in Ref.~\cite{smm1}, by matching the
microscopic results obtained in Ref.~\cite{Brack85}.
Very light
fragments, $A < 5$, are considered as point particles and therefore
contribute to $f_i$ only through their Coulomb and binding energies,
where the latter are taken from empirical data, as well as their spin
degeneracy factors $g_i$. The exception is the alpha particle, in
which case, one suppresses the surface contribution to the excited
states. For all heavier nuclei, $A\ge 5$, $B_i$ is computed through a
given mass formula and it is assumed that $g_i=1$ as the contribution
to the statistical weight is, to some extent, taken into account by
the excitation part of the free energy.

The chemical potentials $\mu_Q$ and $\mu_B$ are obtained by imposing the
constraints given by Eqs.\ (\ref{eq:constA}) and (\ref{eq:constZ}).
In the numerical calculations, the relative error on the mass or proton
numbers is required to be better than $10^{-10}$ in each case.

\subsection{\label{sec:be}The binding energy}
The binding energy of nuclei plays an important role in the determination
of the relative yields produced in the breakup of a nuclear system
\cite{ISMMmass} and therefore it affects the isoscaling parameters.
To investigate this point, we consider three different mass formulae based on
the Liquid Drop Model (LDM) \cite{massFormula}:

\begin{equation}
B_{A,Z}=C_v A - C_s A^{2/3}-C_c\frac{Z^2}{A^{1/3}}+C_d\frac{Z^2}{A}
+\delta_{A,Z}A^{-1/2}
\label{eq:ldm12}
\end{equation}

\noindent
where

\begin{equation}
C_{\rm c}=\frac{a_{\rm c}}{1+\Delta}\;,
\end{equation}

\begin{equation}
C_i=a_i\left[1-k_i\left(\frac{A-2Z}{A}\right)^2\right]
\label{eq:cs}
\end{equation}

\noindent
and $i=v,s$ corresponds to volume and surface terms, respectively. Notice
that $C_v$ and $C_s$ might include contributions associated with the
symmetry energy, as we discuss below. The following two terms
correspond to the ordinary Coulomb reduction of the binding energy and
its correction due to the surface diffuseness, respectively.
In the first two models we consider below, $\Delta=0$,
so the coefficient $C_{\rm c}=a_{\rm c}$ is constant.
This is not the
case for the last mass formula discussed below. The pairing
coefficient has the usual meaning:

\begin{equation}
\delta_{A,Z}=\left\{
   \begin{array}{rl}
      +C_p, & N\, {\rm and}\, Z\, {\rm even}\\
      0, & A\,\, {\rm odd}\\
      -C_p, & N\, {\rm and}\, Z\, {\rm odd}\\
   \end{array}\right.\;,
\label{eq:pairing}
\end{equation}

\noindent
where $N=A-Z$ denotes the neutron number.

The symmetry energy has contributions from the surface and bulk terms, which
can be easily identified by grouping the corresponding factors
from the formula:

\begin{equation}
E_{\rm sym}=-C_{\rm sym}\frac{(A-2Z)^2}{A}
\label{eq:esym}
\end{equation}

\noindent
where

\begin{equation}
C_{\rm sym}\equiv a_vk_v-\frac{a_sk_s}{A^{1/3}}\;.
\label{eq:cSym}
\end{equation}

\noindent
The simplest form of this mass formula, in which $k_s=0$, $C_d=0$,
has been used in many statistical
calculations \cite{isoscWolfgangBotvina1,Bondorf1995}
and will be labeled henceforth as LDM1, whereas the complete formula,
used in Refs. \cite{ISMMmass,ISMMlong}, is denoted as LDM2.
Since the contribution associated with $C_d$ is small, the main difference
between the two formulae is in the inclusion of surface effects in the
symmetry energy in LDM2.

The third mass formula used in this work, LDM3, was introduced in Ref.\
\cite{PawelMasFormula}, which also considers surface effects in the
symmetry energy, but has a different functional form:

\begin{eqnarray}
B_{A,Z}&=&C_v A - C_s A^{2/3}
-\frac{\alpha^*}{1+\frac{\alpha^*}{\beta^*}A^{-1/3}}
\frac{(A-2Z)^2}{A}\nonumber\\
&-&C_c\frac{Z^2}{A^{1/3}}+\delta_{A,Z}\;.
\label{eq:ldmPawel}
\end{eqnarray}

\noindent
This form reflects the well-known correlation between the bulk
symmetry energy and the dependence of the surface energy on the
isospin asymmetry~\cite{Myers69,Lattimer85,Lattimer91}.
Note that $\alpha^*$ and $\beta^*$ do not
correspond to the isoscaling parameters. We chose these symbols so as
to keep the notation as close as possible to that used in the original
work. The Coulomb parameter $\Delta$, in this case, reads

\begin{equation}
\Delta=\frac{1.90539}{A^{2/3}}-\frac{1}{1+\frac{\beta^*}{\alpha^*}A^{1/3}}
\frac{A-2Z}{6Z}\;.
\label{eq:Delta}
\end{equation}

Since this binding energy formula is more complex than that given by Eq.\
(\ref{eq:ldm12}), the symmetry energy cannot be cast in the same form as
Eq.\ (\ref{eq:esym}).
For large $A$ and small asymmetry, it can be approximated by
Eq.\ (\ref{eq:esym}), where

\begin{equation}
C_{\rm sym}\approx \alpha^*-\frac{\alpha^{*2}/\beta^*}{A^{1/3}}\;,
\label{eq:cSymPawel}
\end{equation}

\noindent
which is identical to the result found above.

As pointed out in Ref.\ \cite{PawelMasFormula}, the inclusion of the
normal surface and surface symmetry energies in the mass formula
should be expected as it simply reflects the dependence of the
nuclear energy functional or equation of state on the nuclear
density. The magnitudes or functional forms of the surface symmetry
energy, however, depend on the density dependence of the symmetry
energy, which is poorly constrained. Investigations of the
temperature dependence of the symmetry energy implicitly assume the
temperature dependence to be a result of the underlying density
dependence of the symmetry energy
\cite{isoscaling2,isoYennello2003,isoSouliotis2003}. To model a
temperature dependence without considering the surface symmetry
energies of nuclei is intrinsically inconsistent.

\begin{table*}
\caption{\label{tab:ldmpars}Parameters of the liquid drop mass formulae used in
this work.
All the values are given in MeV.}
\begin{ruledtabular}
\begin{tabular}{cccccccccc}
Label & $a_v$ & $a_s$ & $a_c$ & $C_d$ & $C_p$ & $a_v k_v$ & $a_s k_s$ &
$\alpha^*$ & $\beta^*$ \\
LDM1 & 15.2692 & 16.038 & 0.68698 & 0.0 & 11.277 & 22.3918 & 0.0 & na & na\\
LDM2 & 15.6658 & 18.9952 & 0.72053 & 1.74859 & 10.857 & 27.7976 & 33.7053 &
na & na\\
LDM3 & 15.5586 & 18.2369 & 0.70521 & na & 11.012 & na & na & 27.1508 &
16.6412 \\
\end{tabular}
\end{ruledtabular}
\end{table*}

\

\begin{figure}[tb]
\includegraphics[height=8.0cm,angle=-90]{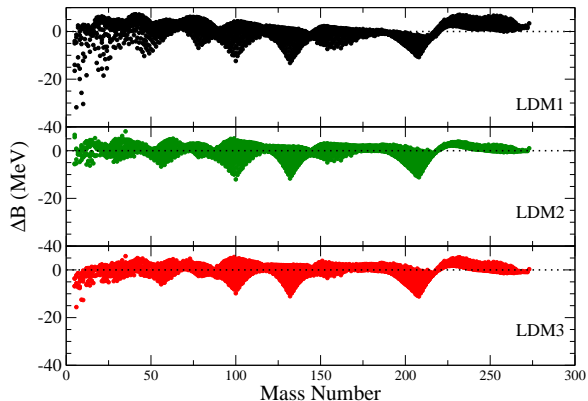}
\caption{\label{fig:be} (Color online) Difference between the calculated
binding energies and the empirical values of Ref. \cite{AudiWapstra} as a
function of the mass number for the three mass formulae used in this work.}
\end{figure}

The parameters of the three mass formulae have been fitted to the
empirical binding energies given in Ref.\ \cite{AudiWapstra}, for
$A>4$. In fig.\ \ref{fig:be}, we show the difference $\Delta B$
between the calculated $B_{A,Z}$ and the empirical values. In all
the cases, one observes a fairly good overall agreement with the
data, except in the regions where shell effects are relevant and
cannot be taken into account by these simple formulae. It is
important to emphasize that $\Delta B$ denotes the total difference
and that it is not divided by the mass number, as this quantity is
usually displayed. Larger discrepancies are observed for light
nuclei, $A<25$, with the LDM1 than for LDM2 or LDM3, because surface
effects in the symmetry energy, which are more important in the
light fragments, is disregarded in the LDM1 mass formula. The other
mass formulae, LDM2 and LDM3, are both more accurate than LDM1, and
LDM2 is slightly more accurate than LDM3.

\

The best fit parameters are listed in table \ref{tab:ldmpars}.
Although paring is switched off in the calculations presented in the next
section, we include it here for completeness.
One may notice that the common parameters in LDM2 and LDM3 are very similar.
The best fit value of $\alpha^*$ is very close to $a_vk_v$, which
explains the resemblance between the predictions of the two mass formulae.
Deviations should be expected for light masses,
where the factor proportional to $1/A^{1/3}$ in Eqs.\ (\ref{eq:cSym}) and
(\ref{eq:cSymPawel}) becomes non-negligible, because
$\alpha^{*2}/\beta^*\approx 44.3$~MeV differs appreciably from
$a_sk_s\approx 33.7$~MeV.

\section{\label{sec:results}Results and discussion\protect}
\begin{figure}[htb]
\includegraphics[height=8.5cm,angle=-90]{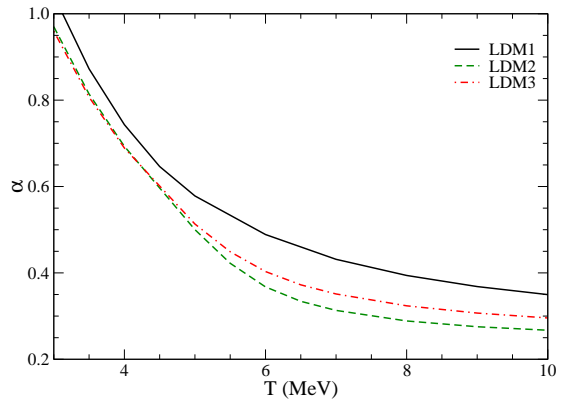}
\caption{\label{fig:alphaLDMs}(Color online) Isoscaling parameter $\alpha$
obtained using different binding energies in the grand canonical calculations
as a function of the temperature.}
\end{figure}

The grand canonical model presented above is now applied to the description
of the breakup of the $^{112}$Sn and the $^{124}$Sn nuclei at
$\rho/\rho_0=1/3$.
In Fig.\ \ref{fig:alphaLDMs}, we show the isoscaling parameter $\alpha$,
obtained using the different mass formulae, as a function of the temperature.
At T=6 MeV, up to 30\% reduction of the isoscaling parameter $\alpha$
can be obtained by including the surface effects in the symmetry energy
(LDM2/LDM3) (see Fig.\ \ref{fig:alphaLDMs}), while preserving the good
agreement with the empirical binding energies.
The final values for $\alpha$ are comparable to the data of ref.\cite{isoscWolfgangBotvina1}, which suggest that one may be able to reproduce data by adopting a
more accurate mass formula without grossly changing the symmetry energy values
obtained from the fitting of the nuclear binding energies. Even so, the results reveal that distinct parameterizations of the surface symmetry energy energy do
lead to appreciably different values of $\alpha$, in spite of the fact that
the mass formulae are, to a large extent, equally accurate. So the parameterizations adopted here may still not be the optimal ones.

This differences in the predictions for $\alpha$ for the different mass models may be understood from the relationship between $\alpha$
and the symmetry energy in Eq.\ (\ref{eq:alphaGamma}), as discussed
in Refs.\ \cite{isoscaling2,isoscWolfgangBotvina1}. At low
temperatures, the chemical potentials are strongly influenced by the
heavy fragments, which dominate the mass distribution. Since, in
this mass region, the coefficient $C_{\rm sym}$ in LDM2 and in LDM3
is nearly the same, as just discussed, Eq.\ (\ref{eq:alphaGamma})
predicts that the isoscaling parameter $\alpha$ obtained with these
two mass formulae should be very similar. The opposite holds in the
case of the LDM1. One should note that Eq.\ (\ref{eq:alphaGamma})
has been derived at very low temperatures
\cite{isoscWolfgangBotvina1}. As the temperature increases, more
light fragments with $A<50$ are produced and the changes in the
chemical potentials in the $^{112}$Sn and $^{124}$Sn systems are
appreciably affected by them. Therefore, the binding energy
differences observed in Fig.\ \ref{fig:be}, in this lighter mass
region, become very important. This explains the increase of the
discrepancy between the values of $\alpha$ obtained using LDM2 and
LDM3 at high temperatures. Eventually, our three curves will
converge again at much higher temperatures ($T>>10$~MeV), where only
very light particles are produced. This latter convergence would
occur because we use empirical binding energies in this mass region
in the statistical calculations in all the cases.

The difference between the values of $\alpha$ obtained using LDM1
and LDM2/LDM3 is important for interpreting experimental data
because in recent studies the symmetry energy coefficient has been
adjusted in order to reproduce the experimental values of $\alpha$
\cite{isoscalingIndraGSI2005}. Our results indicate that a large
temperature dependence to the symmetry energy may not be necessary
if a suitable parametrization of the binding energies are
incorporated in the models. Nevertheless, we continue below
to discuss another issues that may be important to consider
if it still appears interesting to include an additional temperature
dependence to the symmetry energy.

\begin{figure}[tb]
\includegraphics[height=8.0cm]{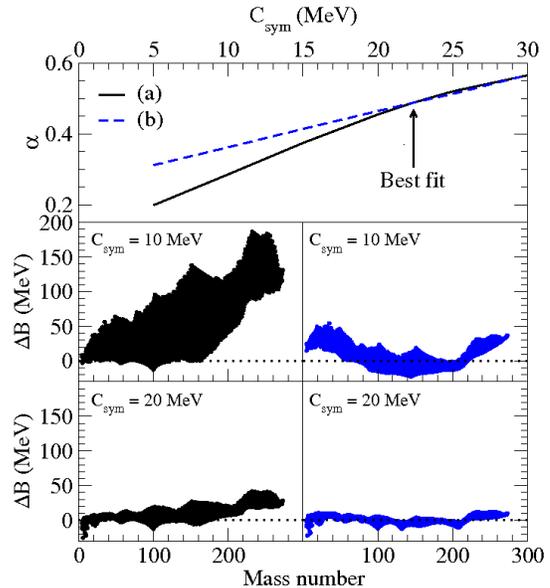}
\caption{\label{fig:alphaCsym} (Color online) Upper panel: isoscaling parameter
$\alpha$ obtained using LDM1 as a function of $C_{\rm sym}$.
Lower panels: $\Delta B$ for different values of $C_{\rm sym}$.
For details, see the text.}
\end{figure}

We now examine the consequences of changing $C_{\rm sym}$ in
statistical calculations, as has been made in different studies that
focused on the isoscaling analysis
\cite{isoscalingIndraGSI2005,isoYenello2005,isoSymmetryBotvina2006,symEnergyShatty2007}
and on the isotopic distribution
\cite{IsospinSymmetry,BotvinaSurf2006}. The top panel of Fig.\
\ref{fig:alphaCsym} displays the parameter $\alpha$ obtained using
LDM1 for different values of $C_{\rm sym}=a_vk_v$, for $T=6$~MeV.
The full line represents the results obtained in the approach of
\cite{isoscalingIndraGSI2005,isoYenello2005,isoSymmetryBotvina2006,symEnergyShatty2007}
and \cite{IsospinSymmetry,BotvinaSurf2006}, case (a), where all the
parameters of the mass formula are kept fixed, except for $k_s$. The
values for $\alpha$ decrease as the $C_{\rm sym}$ decreases, as
noted by Refs.
\cite{isoscalingIndraGSI2005,isoYenello2005,isoSymmetryBotvina2006,symEnergyShatty2007}.
The left side of the lower two panels in Fig. 3 display the
difference $\Delta B$ between the calculated $B_{A,Z}$ and the
empirical values. Reducing the symmetry energy coefficient to
$C_{\rm sym}=10$~MeV, similar to the values deduced in Ref. \cite{isoscalingIndraGSI2005}, increases the binding of the most neutron-rich fragments of Mass number 100 by up to 100 MeV without changing the binding of the symmetric fragments. If the authors of Refs.
\cite{isoscalingIndraGSI2005,isoYenello2005,isoSymmetryBotvina2006,symEnergyShatty2007}
are correct, why does it occur? Without changing the nuclear density, such
a large effect can only be due to changes in the interaction component of
the symmetry energy; recent calculations of the temperature
dependence for interaction component of the symmetry energy predicts a 
a reduction in $C_{\rm sym}$ of less than 5~MeV \cite{fabio1,fabio2,tempsymenergyli2008}. 

Indeed, the authors of
\cite{isoYenello2005,isoSymmetryBotvina2006,symEnergyShatty2007},
explain their reduction in $C_{\rm sym}$ to be the result of a
reduction in nuclear density. This plausible
explanation for a reduced value for $C_{\rm sym}$ does not accurately
represent the calculations that were performed by Refs.\
\cite{isoYenello2005,isoSymmetryBotvina2006,symEnergyShatty2007}, however.
While it is true 
that the overall density of the multi-fragmenting systems studied by Refs.\
\cite{isoYenello2005,isoSymmetryBotvina2006,symEnergyShatty2007} is reduced,
according to the descriptions given in these studies and in the
references they cite, the increased volume is mainly due to the separation
between the fragments.
The fragments in their SMM calculations have the bulk, surface and Coulomb energies of nuclei at normal density, and the excluded volume corrections in their calculations assume all of
the fragments to be at normal density. 

We note that it is possible to reduce the overbinding for the
heaviest neutron-rich isotopes by refitting the binding energy
expression used in the free energy calculation of hot nuclei to the
measured masses. We have done this and show the results as case (b)
in the right panels of Fig. 3. The dashed line in the upper panel of
this figure shows the dependence of $\alpha$ on $C_{\rm sym}$ to be
rather similar to that for case (a). After refitting, the
neutron-rich nuclei with $A<30$ are overbound by values ranging up to
about 2 MeV/nucleon, indicating a considerable enhancement in the
attractive potential energy contribution for such neutron-rich
nuclei.

The results for both case (a) and case (b) clearly show that
$\alpha$ is strongly sensitive to $C_{\rm sym}$, as reported
previously \cite{isoscalingIndraGSI2005}. For $C_{\rm sym}>22$~MeV,
there is almost no difference whether one changes the other
parameters of the mass formula, since the symmetry energy becomes so
important that its contribution to the determination of $\alpha$
overwhelms the other terms. On the other hand, for smaller $C_{\rm
sym}$ values, the other terms of the mass formula, neglected in the
derivation of Eq.\ (\ref{eq:alphaGamma}), become non-negligible and
therefore the two procedures lead to distinct results. 

For $C_{\rm sym}$ below the best fit value, values for $\Delta B$ 
of the order of 2 MeV/nucleon are obtained for light nuclei $A<20$ at
$C_{\rm sym}=10$~MeV. We note that such light nuclei are the dominant species 
at high temperatures in any SMM calculation, whether for case (a), (b) or for
the calculations with LDM1, LDM2 or LDM3 shown earlier. Thus, all attempts within the
SMM approach to explain the values of $\alpha$ reported by 
\cite{isoscalingIndraGSI2005,isoYenello2005,isoSymmetryBotvina2006,symEnergyShatty2007}
involve modifying the free energies of the light fragments with $A<30$. The most important theoretically justified modification to these light fragment free energies is to include surface symmetry energy terms, to reflect the strong influence of the nuclear surface on the symmetry energies of light nuclei. The option of reducing $C_{\rm sym}$ to low values like $C_{\rm sym}=10$~MeV, while leading to similar reductions in $\alpha$, does not seem to have a clear theoretical
justification.
More theoretical work is needed to explain why it is reasonable to expect such
a large reduction in $C_{\rm sym}$ and to justify the particular form of the
reduction chosen in 
Refs. \cite{isoscalingIndraGSI2005,isoYenello2005,isoSymmetryBotvina2006,symEnergyShatty2007}.

\begin{figure}[tb]
\includegraphics[height=8.5cm,angle=-90]{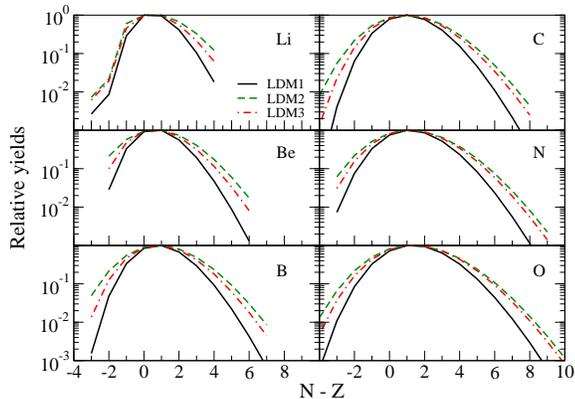}
\caption{\label{fig:isoDist}(Color online) Isotopic distribution of fragments
produced in the decay of the $^{112}$Sn nucleus at $T=6$~MeV.}
\end{figure}

We note that $C_{\rm sym}$ has also been reduced in different studies
\cite{IsospinSymmetry,BotvinaSurf2006} in order to obtain broader
isotopic distributions. The inclusion of surface symmetry energy terms also
has a similar effect in this case. 
We show, in Fig.\ \ref{fig:isoDist}, the isotopic distribution for
some light nuclei, calculated for the breakup of the $^{112}$Sn at
$T=6$~MeV, using the three different mass formulae discussed in this
work, with the best fit parameters listed in table \ref{tab:ldmpars}.
These results clearly show that the $A$ dependence of $C_{\rm sym}$ in
LDM2 and LDM3 leads to wider distributions as $C_{\rm sym}$ becomes smaller
for light nuclei, due to the influence of the symmetry energy.

\section{\label{sec:conclusions}Concluding remarks\protect}

We have analyzed the role played by the mass formula used in
statistical calculations to determine the isoscaling parameter
$\alpha$, as well as in the isotopic distribution of fragments
produced in the breakup of a thermally excited nucleus.
Our results show that the reduction of the symmetry energy values
\cite{isoscalingIndraGSI2005,isoYenello2005,isoSymmetryBotvina2006,symEnergyShatty2007,IsospinSymmetry,BotvinaSurf2006},
which best fit the empirical masses, may not be necessary
to reproduce the experimental $\alpha$ and the isotopic distributions.
Rather, the observed experimental reduction of the isoscaling parameter
\cite{isoscalingIndraGSI2005,isoSymmetryBotvina2006,symEnergyShatty2007} might
be reproduced if the surface corrections to the symmetry energy are included in
the mass parametrization used in the statistical multifragmentation models.
Since non-negligible deviations from the optimal
value of $C_{\rm sym}$ lead to large discrepancies between the
theoretical and the empirical binding energies  and free energies used in the model
calculations, such changes should not be taken lightly especially if
similar effects can be reproduced by the inclusion of surface effects
in the mass formula. Certainly, some additional theoretical work is needed to justify the forms of the modifications to the symmetry energies that have been chosen in Refs. \cite{isoscalingIndraGSI2005,isoSymmetryBotvina2006,symEnergyShatty2007}. Furthermore, as pointed out in Ref.\
\cite{nuclearThermometry2000}, improved treatments to the decay of the
primary fragments than the usual Weisskopf approach
\cite{grandCanonicalBotvina1987,flowSMM} also contribute to isotopic
distributions with larger tails.
Finally, we believe that our conclusions will
not be affected by more realistic treatments which may introduce a
temperature dependence to the symmetry energy since the corresponding
effects, at a fixed density, are expected to be small
\cite{fabio1,fabio2,tempsymenergyli2008}.

\begin{acknowledgments}
We would like to acknowledge CNPq, FAPERJ, and the PRONEX
program under contract No E-26/171.528/2006,
for partial financial support.
This work was supported in part by the National Science Foundation under Grant
Nos.\ PHY-0606007 and INT-0228058.
AWS is supported by Joint Institute for Nuclear Astrophysics at MSU
under NSF-PFC grant PHY 02-16783.
\end{acknowledgments}

\bibliography{isoMasFormula}

\end{document}